\documentclass[12pt]{spieman}  
\usepackage{amsmath,amsfonts,amssymb,bm}
\usepackage{graphicx, svg}
\usepackage{subcaption}
\usepackage{setspace}
\usepackage{tocloft}
\usepackage{siunitx}  
\usepackage{overpic}

\newcommand{\bb}{\mathbf{b}}

\newcommand{\bx}{\mathbf{x}}

\newcommand{\bA}{\mathbf{A}}

\newcommand{\bX}{\mathbf{X}}

\newcommand{\bPhi}{\boldsymbol{\Phi}}

\DeclareMathOperator*{\argmin}{arg\rm{}min}  
\newcommand{\Az}{\mathrm{Az}}
\newcommand{\El}{\mathrm{El}}

\newcommand{\REV}[1]{{\color{black} #1}} 

\usepackage{lineno} 

\title{Dynamic Mode Decomposition for Aero-Optic Wavefront Characterization}

\author[a,$\dagger$]{\REV{Shervin Sahba}}
\author[b,$\dagger$]{\REV{Diya Sashidhar}}
\author[c]{Christopher C. Wilcox}
\author[c]{Austin McDaniel}
\author[d]{Steven L. Brunton}
\author[a,*]{J. Nathan Kutz}
\affil[a]{Department of Physics, University of Washington, Seattle, WA 98195}
\affil[b]{Department of Applied Mathematics, University of Washington, Seattle, WA 98195}
\affil[c]{US Air Force Research Laboratory, 3550 Aberdeen Ave SW, Kirtland AFB, NM 87117}
\affil[d]{Department of Mechanical Engineering, University of Washington, Seattle, WA 98195}

\cftpagenumbersoff{figure}
\cftpagenumbersoff{table} 
\begin{document} 
\maketitle

\begin{abstract}
Aero-optical beam control relies on the development of low-latency forecasting techniques to quickly predict wavefronts aberrated by the Turbulent Boundary Layer (TBL) around an airborne optical system\REV{, and its study applies to a multi-domain need from astronomy to microscopy for high-fidelity laser propagation}. We leverage the forecasting capabilities of the Dynamic Mode Decomposition (DMD) -- an equation-free, data-driven method for identifying coherent flow structures and their associated spatiotemporal dynamics -- in order to estimate future state wavefront phase aberrations to feed into an adaptive optic (AO) control loop. We specifically leverage the optimized DMD (opt-DMD) algorithm on a subset of the Airborne Aero-Optics Laboratory Transonic (AAOL-T) experimental dataset, \REV{characterizing aberrated wavefront dynamics for 23 beam propagation directions via the spatiotemporal decomposition underlying DMD}. Critically, \REV{we show that opt-DMD produces an optimally de-biased eigenvalue spectrum with imaginary eigenvalues, allowing for arbitrarily long forecasting to produce a robust future-state prediction, while exact DMD loses structural information due to modal decay rates.}
\end{abstract}


{\noindent \footnotesize\textbf{*}J. Nathan Kutz,  \linkable{kutz@uw.edu} }

{\noindent \footnotesize\textbf{$\dagger$}These authors contributed equally to this work.}

\begin{spacing}{1}   

\section{Introduction}\label{sect:intro} 

\REV{Free-space communication, high-resolution imaging, and directed energy are sought-after lasing applications, all of which are desirable on aircraft. Achieving high-fidelity laser beam propagation in the air requires mitigating the phase distortion of the outgoing wavefront. In atmospheric and free-space laser propagation, such as in observational astronomy~\cite{daviesAdaptiveOpticsAstronomy2012} and satellite quantum key distribution~\cite{caoLongdistanceFreespaceMeasurementdeviceindependent2020}, and even in biological specimen microscopy~\cite{boothAdaptiveOpticalMicroscopy2014}, spatiotemporal variations of the index of refraction must be compensated for via adaptive optic (AO) control systems. These AO systems typically use deformable mirrors to correct the outgoing beam by preemptively deforming the wavefront to cancel out any subsequent perturbation. Developing robust and responsive predictive controllers for AO systems is a highly desired enhancement with application well beyond airborne optics.}

For an airborne optical platform, the AO system must correct wavefront distortions resulting from three primary sources: mechanical jitter of the platform, near-field effects where the Turbulent Boundary Layer (TBL) around the airborne platform rapidly alters the refractive index, and atmospheric effects where inhomogeneities and turbulence alter the propagation medium.~\cite{Jumper2017, deluccaEffectsEngineAcoustic2018} 
This paper focuses on the near-field wavefront distortions that are referred to as aero-optical effects. 

\begin{figure}[t]
\begin{center}
\begin{tabular}{c}
\includegraphics[width=0.96\columnwidth]{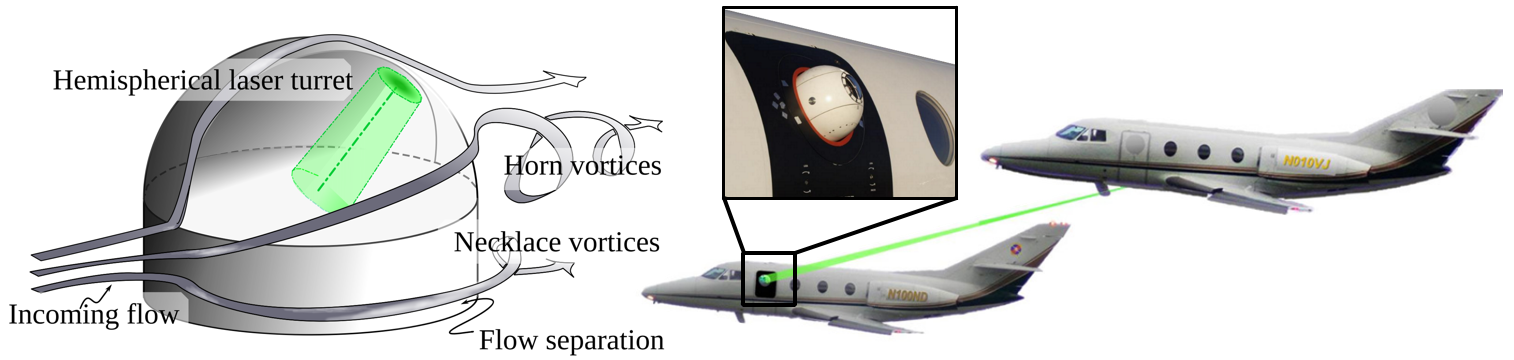}
\end{tabular}
\end{center}
\caption{AAOL-T aircraft with hemispherical laser turret.  The turret geometry produces a turbulent flow field in the few centimeters surrounding the sensors. \REV{Because the flow dynamics may not be accounted for with on-board sensors across applications, their effects must be predictively controlled in order to properly produce high-fidelity, coherent transmission of the electric field.} }\label{fig:overview} 
\end{figure}

The term aero-optics refers to the intersection of optical and aerodynamic phenomena, such as the effects on the optical field from a high-speed turbulent flow, where air is forced \REV{around} the optical system, possibly resulting in flow separation and shock formation. Characterizing these rapid wavefront aberrations is the goal of this study\REV{, using experimental data from the Airborne Aero-Optics Laboratory Transonic (AAOL-T)~\cite{jumperAirborneAeroOpticsLaboratory2015}. As depicted in Figure \ref{fig:overview}, the AAOL-T consists of a pair of Falcon 10 aircraft measuring in-flight laser transmission. The aero-optic effects that determine wavefront deformation are dependent on such factors as the laser platform shape and aircraft geometry, the speed of the craft, the Reynolds number, and the direction of beam propagation, which will be further discussed in Section \ref{sec:aaolt}.}

The \REV{aerodynamic environment of airborne laser platforms motivates} high-fidelity computational fluid dynamics (CFD) models and experimental techniques in the study of aero-optics effects. The high-speed, high Reynolds number compressible flows around airborne platforms can contain TBLs, shear layers, and wakes, as well as shock waves in the case of transonic and supersonic flows.~\cite{wangPhysicsComputationAeroOptics2012, Jumper2017} As a laser beam propagates through this turbulent flow surrounding the aperture, refractive index fluctuations cause phase aberrations, and the resulting distortions of the optical field are referred to as aero-optical effects~\cite{Wilcox2020}. 

The index of refraction, $n$, is directly linked to air density fluctuations by
\begin{equation}
    n({\bf r}) = 1 + K_{GD}(\lambda_0)\rho({\bf r}),
    \label{Eq:GladstoneDaleDefinition}
\end{equation}
where $K_{GD}$ is the wavelength-dependent Gladstone-Dale factor, $\lambda_0$ is the laser wavelength, and $\rho({\bf r})$ is the air density as a function of the spatial variable ${\bf r}$~\cite{wangPhysicsComputationAeroOptics2012}. 

\begin{figure}[t]
\begin{center}
\begin{tabular}{c}
\includegraphics[width=0.9\columnwidth]{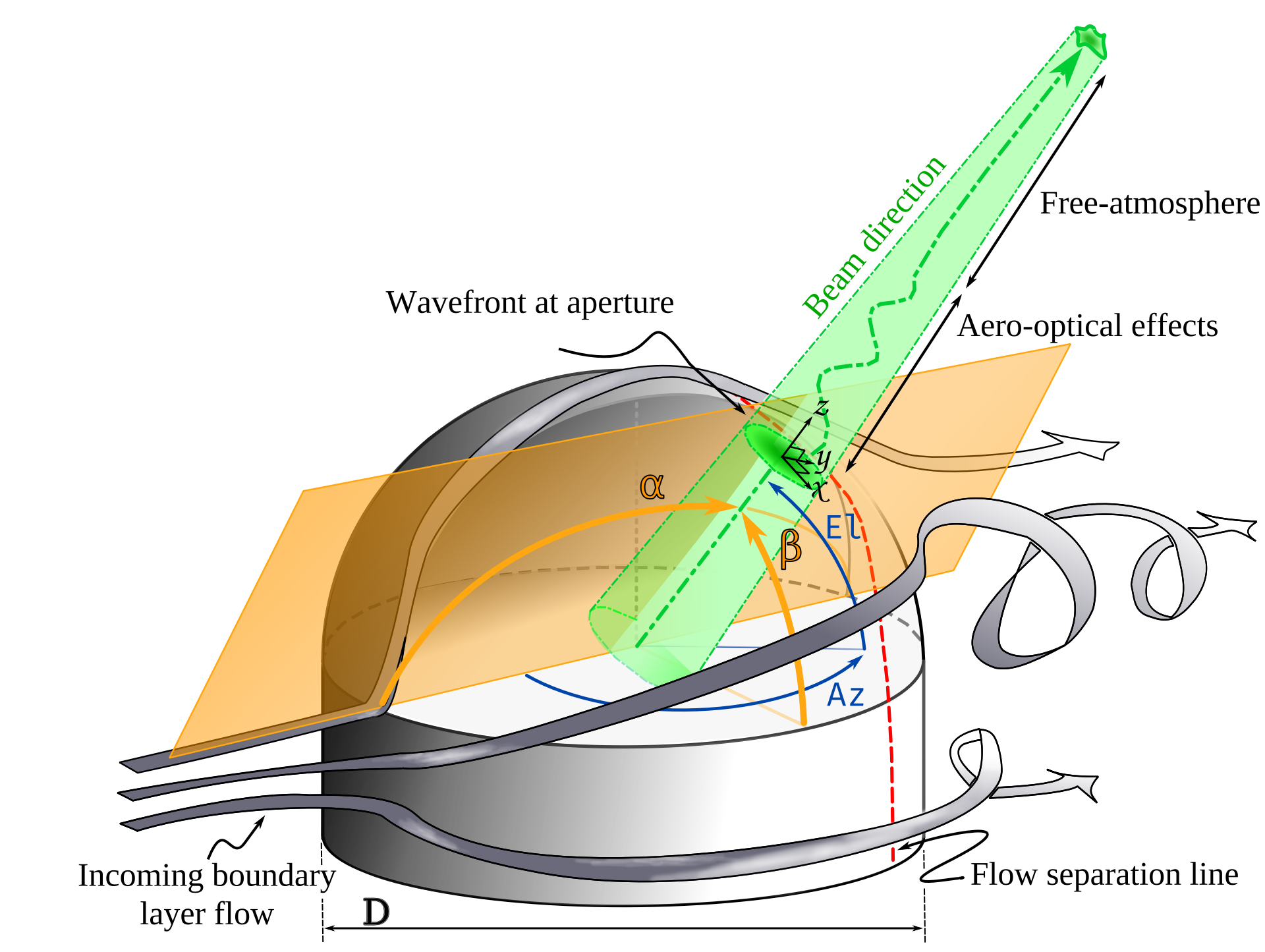}
\end{tabular}
\end{center}
\caption 
{ \label{fig:turret} Detail of the turret geometry.  \REV{The aperture (green disc with local coordinate system $x$-$y$-$z$) images the aberrated wavefronts. Wavefront distortion is highly dependent on beam direction, parametrized by $\alpha$ and $\beta$, the look-back angle and inclination respectively. These two parameters are mapped from the azimuth and elevation angles, $\Az$ and $\El$, of the cylindrical base using the transformation in Equation \ref{eqn:alphabeta}}. The \REV{red dotted line} indicates the beginnings of the flow separation region for $\alpha>\pi/2$ and turbulent coherent structure formation.}  
\end{figure}

Characterizing propagating beam wavefront dynamics in the TBL is critical to correcting the outgoing phase profile of the beam. From an applied standpoint, despite the relatively short\REV{, centimeter scale,} distance traveled in the TBL, the beam quality is immediately and often heavily degraded within this region.~\cite{wang2012cmame} A \REV{typical method} to quantify the aero-optic wavefront aberrations from a given refractive index field is by calculating optical path difference (OPD). OPD is computed by first calculating the optical path length (OPL), which is proportional to the travel time for corresponding rays. OPL is often computed as the integral of the index of refraction along the propagation direction,
\begin{equation}
    \text{OPL}(x,y,t) = \int_0^{z_1} n(x,y,z,t) dz.
    \label{eqn:OPL-definition}
\end{equation}
Subtracting the mean OPL over the spatial coordinates of the aperture produces the OPD,
\begin{equation}
    \text{OPD}(x,y, t) =\text{OPL}(x,y,t) - \langle \text{OPL}(x,y, t) \rangle.
    \label{eqn:OPD-definition}
\end{equation}
We have let $z$ \REV{in Equation} \ref{eqn:OPL-definition} be the optical axis of the beam with $x$ and $y$ coordinates covering the aperture as seen in Figure \ref{fig:turret}. Assuming the dominant contribution to the OPD occurs within the TBL over short transmission distances, we may let the upper bound of integration, $z_1$, match the extent of the TBL. The root mean square of OPD across each dataset provides a metric to assess the severity of wavefront distortions for the given experiment. To compare OPD across experiments, we then normalize it as a dimensionless quantity, \REV{assigning}
\begin{equation} \label{eqn:normalizedOPD}
    \REV{\text{OPD} \leftarrow \frac{ \text{OPD}}{M^2D \rho/\rho_0},}
\end{equation}
where $M$ is the Mach number as a ratio of the speed of sound, $D$  is the turret diameter as seen in Figure \ref{fig:turret}, and $\rho/\rho_0$ is a ratio of in-flight air density to sea-level air density. For the AAOL-T data, we analyze a subset such that each trial is taken at $M=0.6$ and $\rho = \SI{0.812}{\kilo\gram/\meter^3}$. The sea level air density is set to the standard $\rho_0 = \SI{1.225}{\kilo\gram/\meter^3}$, and the AAOL-T turret diameter is $D = \SI{0.3048}{\meter}$. \REV{To characterize each trial in our dataset, we compute the root-mean-square, $\text{OPD}_{\text{rms}}$. This  scaling allows extrapolation across various turret diameters and altitudes, since the same spatial wavefront characteristics are retained for subsonic flow ($M\lessapprox0.6$) in general and for transonic ($M\gtrapprox0.6$) and supersonic ($M > 1$) flow as long as Mach number is matched~\cite{jumperAirborneAeroOpticsLaboratory2015}.} When referring to $\text{OPD}$ and $\text{OPD}_{\text{rms}}$ in figures and elsewhere in this paper, we imply the normalized formulation \REV{in Equation \ref{eqn:normalizedOPD}.}

The analysis of aero-optical wavefront reconstruction leverages time-series measurements collected through the TBL. Here we highlight the measurement and sensor technologies \REV{used} for characterizing aero-optic interactions. \REV{We describe} the underlying mathematical architecture that leverages these measurements in order to \REV{develop} dynamic models for wavefront reconstruction. 

\section{\REV{AAOL-T Experimental Data}} \label{sec:aaolt}
The AAOL-T was run by researchers at the University of Notre Dame to obtain live aero-optical data in flight. A $\SI{532}{\nano\meter}$ source beam propagates from a hemispherical laser turret of diameter $\SI{0.3048}{\meter}$ mounted on a Falcon 10 aircraft, as depicted in Figure \ref{fig:overview}. The beam overfills the pupil aperture on the receiver laboratory aircraft. A Shack-Hartmann wavefront sensor (SHWFS), depicted in Figure \ref{fig:SH-diagram} and described in Section \ref{sec:shwfs}, is used to capture wavefront phase aberrations between the source and destination beams.

The AAOL-T dataset involves measurements up to Mach 0.8 \REV{taken at a sampling rate of 25 kHz.} \REV{Shock-formation on the hemispherical turret is observed at transonic Mach numbers~\cite{gordeyevFluidDynamicsAerooptics2010}.} The distance between source and receiver aircraft is approximately 50 m. The beam direction is recorded in terms of its azimuth and elevation angles, as visualized in Figure \ref{fig:turret}, with respect to \REV{the cylindrical base of the hemispherical turret. It is useful to reparametrize the beam direction} in terms of a “look-back” angle and inclination angle, $\alpha$ and $\beta$ respectively, where

\begin{equation} \label{eqn:alphabeta}
    \alpha = \cos^{-1}\left[\cos(\Az)\cos(\El)\right]
\end{equation}
\begin{equation}
    \beta = \tan^{-1}\left[\frac{\tan(\El)}{\sin(\Az)}\right].
\end{equation}

\begin{figure}
     \centering
     \begin{overpic}
     [width = 1\textwidth]{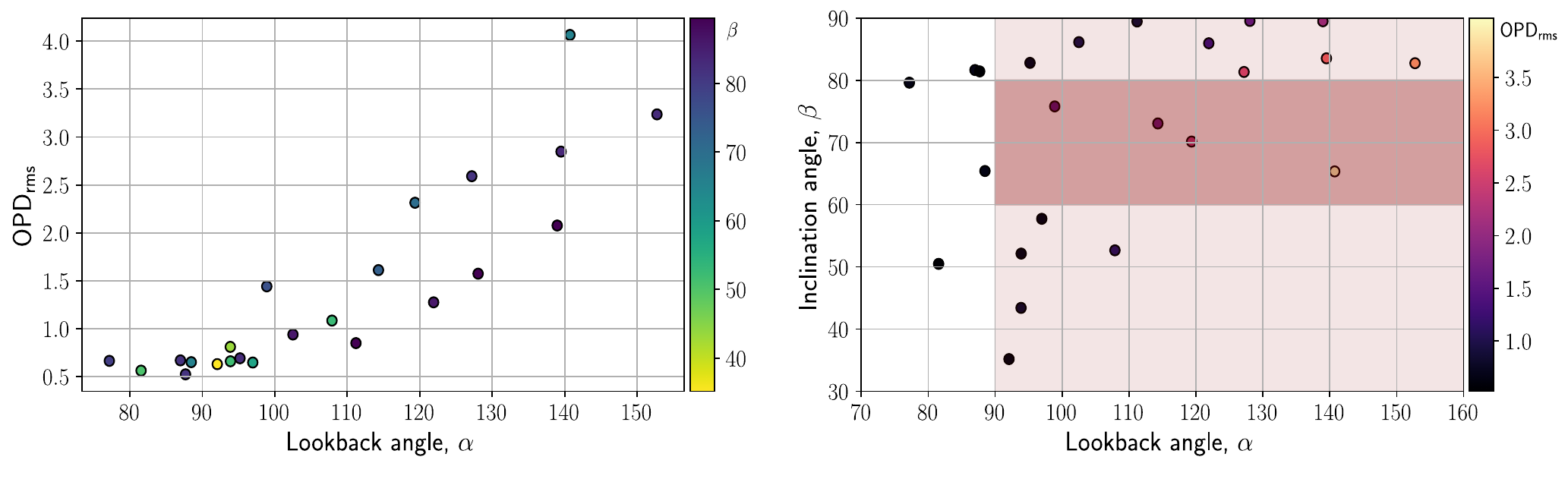}
     \put(-1,5){(a)} \put(49,5){(b)}
     \end{overpic}
     {\phantomsubcaption\label{fig:AlphaBetaOverview-a}}
     {\phantomsubcaption\label{fig:AlphaBetaOverview-b}}
     \caption{\REV{Beam propagation direction affects OPD, as seen in the AAOL-T data.} \textbf{(a)} \REV{The effect of look-back angle, $\alpha$, on OPD$_{\text{rms}}$ with inclination angle, $\beta$ given by the color bar.}  OPD increases as the turret direction $\alpha$ looks to the trailing edge of the flow field. \textbf{(b)} \REV{Look-back angle versus inclination angle with OPD$_{\text{rms}}$ given by the color bar. The light red shaded region denotes look-back angles $\alpha>90^\circ$. Beam directions in this region may be in the flow separation region of the index of refraction. The dark red shaded region denotes inclination angles $60^\circ < \beta < 80^\circ$. This coloring serves as a rough guide to where horn vortices interrupt the boundary layer flow, resulting in greater OPD.} \label{fig:AlphaBetaOverview}}
\end{figure}

\REV{A look-back angle, $\alpha$, of zero is a beam propagating in the forward direction of the aircraft, while an angle of $180^\circ$ would designate propagation towards the rear, through the outgoing turbulent wake. An inclination angle, $\beta$, of zero describes a beam facing the earth, while $180^\circ$ would be skyward.} Figure \ref{fig:AlphaBetaOverview-a} measures the effects of $\alpha$ and $\beta$ on OPD$_\text{rms}$ for all 23 data sets. For backwards looking angles $\alpha>90^\circ$, an increasing OPD indicates a heightening level of aberrations in the wavefront. With Figure \ref{fig:AlphaBetaOverview-b}, we can visualize the effects of this look-back angle when also considering $\beta$. The dark red region indicates angles that lie where horn vortices exist; OPD${_\text{rms}}$ tends to be greatest for these data points.

In-flight measurements from the AAOL-T experiment, the aircraft, and hemispherical laser turret with conformal window, which are depicted in Figures \ref{fig:overview} and \ref{fig:turret}, have contributed to a database of aero-optic disturbance measurements. Atmospheric aberrations are often characterized by Zernike modes~\cite{tangoCirclePolynomialsZernike1977}, an orthogonal sequence of polynomials that span the unit disk and possess odd or even radial symmetries. Zernike modes offer interpretability to optical dynamics and can yield insights where radially symmetric aberrations are concerned. Yet this is often not the case for aero-optical disturbances in the TBL which are prone to quickly-varying nonlinearities in the index of refraction at transonic flow speeds~\cite{JosephGoodmanIntroduction}.  An analysis of the temporal phase structure function and other statistics of AAOL-T wavefront data was performed by Brennan and Wittich in 2013~\cite{Brennan2013}. Proper Orthogonal Decomposition (POD) and Dynamic Mode Decomposition (DMD) modes have been used to provide a spatio-temporal characterization of the flow dynamics~\cite{Goorskey2013}. Predictive control methods for aero-optics have been analyzed on these data as well~\cite{goorskeyEfficacyPredictiveWavefront2013, burnsRobustModificationPredictive2016}.

\section{Sensors and Data Acquisition} \label{sec:shwfs}

\begin{figure}[t]
\centering
\begin{overpic}[width = 1\textwidth]{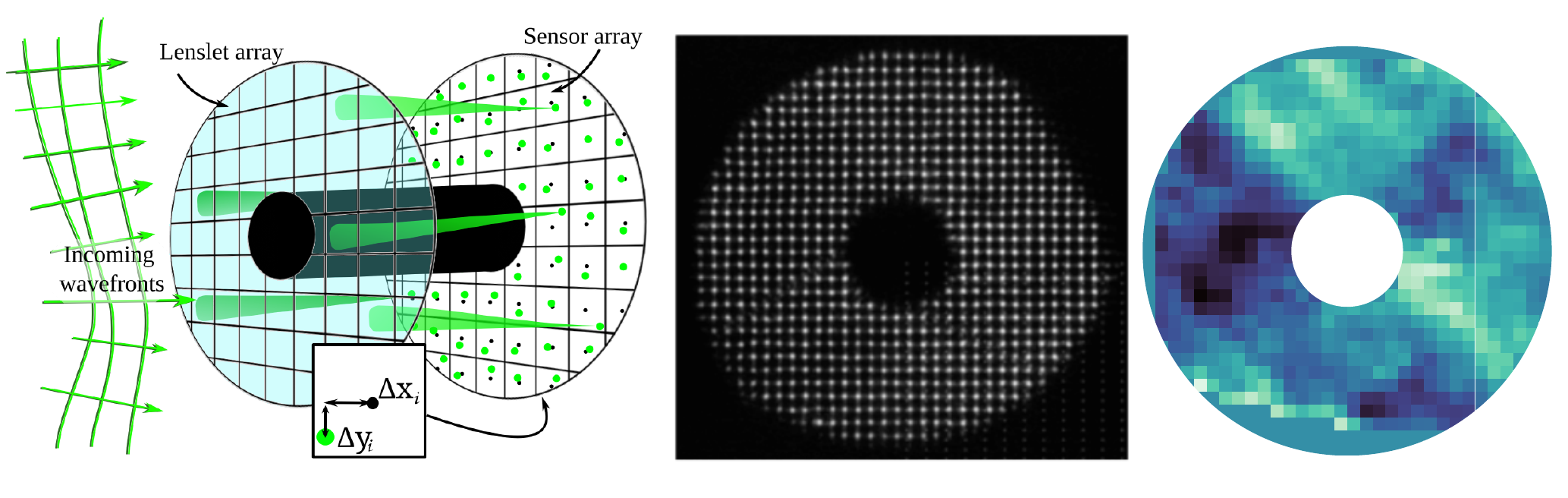}
\put(0,3){(a)} \put(44,3){{\color{white}(b)}} \put(74,3){(c)}
\end{overpic}
{\phantomsubcaption\label{fig:SH-diagram}}
{\phantomsubcaption\label{fig:SH-unprocessed}}
{\phantomsubcaption\label{fig:SH-data}}
\caption{ \textbf{(a)} Geometry of the \REV{SHWFS~\cite{shack1971production}} on the AAOL-T laser turret with an incident aberrated wavefront.  The lenslet arrays project to the sensor array where the displacements from the sensor centroids, measured by $\Delta x_i$ and $\Delta y_i$, is used to compute the local tilts of the wavefront for reconstruction. \textbf{(b)} Unprocessed SHWFS data showing intensities projected on the 32x32 subapertures sensors from the AAOL-T. \textbf{(c)} Processed SHWFS data used in the Dynamic Mode Decomposition (DMD) analysis. \label{fig:SH}} 
\end{figure}

\REV{Depicted in Figure \ref{fig:SH-diagram},} a SHWFS is used to capture wavefront phase aberrations in the AAOL-T experiment.~\cite{shack1971production} A lenslet array in the pupil plane and at a focal distance away from an optical sensor focuses an incoming wavefront into sub-regions on the detector plane. Any deviations from a planar wavefront manifest as displacements, $\Delta x_i$ and $\Delta y_i$, from the optical axis in the $i^{th}$ sub-region. \REV{The wavefront phase can be reconstructed by a least-squares fit of the average intensity-weighted gradients across the subapertures~\cite{shack1971production}. These gradients are proportional to the centroid tilts, the displacement of the centroid of the focused rays. The difference between the centroid tilts and the true gradients can alter SHWFS readings. This source of measurement error cannot be analytically evaluated outside low Mach numbers where weak scintillation and Rytov theory apply~\cite{robertScintillationPhaseAnisoplanatism2006}. A performant SHWFS typically requires the length of each subaperture to be less than a fourth of the Fried coherence length for atmospheric turbulence~\cite{barchersEvaluationPerformanceHartmann2002a}.}

A 3-D representation of the SHWFS on the AAOL-T laser turret is depicted in Figure \ref{fig:SH-diagram}. The incoming beam's aberrated wavefront is focused from the gridded lenslet array onto sub-regions on the detector plane as depicted by the larger green dots. The displacements, $\Delta x_i$ and $\Delta y_i$, of each focused sub-beam from the centroid of each sub-region, shown by the smaller black dot, is used to compute the local tilt of the incoming wavefront, from which the wavefront may be reconstructed. With a planar, unaberrated incoming wavefront, the focused spots would be in a perfect grid matching the lenslet array geometry. The central circular region in Figure \ref{fig:SH-unprocessed} represents the secondary mirror obscuration of the optical system inside the turret, which includes a telescope used to align the turret on one AAOL-T aircraft to the incoming beam from the other and is not an explicit feature of a general SHWFS. Because of \REV{the 1-inch diameter obscuration~\cite{deluccaEffectsEngineAcoustic2018},} the unprocessed SHWFS data is taken as a set of points lying in an annulus.

Figure \ref{fig:SH-unprocessed} shows a single frame of unprocessed SHWFS data from the AAOL-T platform used in this study. The data were acquired using a v1610 Vision Research Phantom camera at 30 kHz for a total of 21,504 frames captured per dataset. Figure \ref{fig:SH-data} is an example of the processed data and the wavefront from the local tilts of the SHWFS that we use in our analysis. As will be described in the upcoming sections, this study investigated 23 sets of data with varying $\alpha$ and $\beta$ angles, as defined in Figure \ref{fig:turret}. \REV{Each $(\alpha,\beta)$ pair defines a beam direction.}

\section{Optimized Dynamic Mode Decomposition}

DMD was an algorithm developed by Schmid~\cite{schmid2008aps,schmid2010jfm} in the fluid dynamics community to identify spatio-temporal coherent structures from high-dimensional data.
DMD is based on \REV{Proper Orthogonal Decomposition (POD)}, which utilizes the computationally efficient singular value decomposition (SVD) so that it scales well to provide effective dimensionality reduction in high-dimensional systems. DMD provides a modal decomposition where each mode consists of spatially correlated structures that have the same linear behavior in time (e.g., oscillations at a given frequency with growth or decay).
Thus, DMD not only provides dimensionality reduction in terms of a reduced set of modes, but also provides a model for how these modes evolve in time.

Several algorithms have been proposed for DMD, with the \emph{exact} DMD framework developed by Tu et al.~\cite{tu2014jcd} being the simplest, least-squares regression to produce the decomposition. DMD is inherently data-driven, and the first step is to collect a number of pairs of snapshots\index{snapshot} of the state of a system as it evolves in time.
These snapshot pairs may be denoted by $\{\bx(t_k),\bx(t_k')\}_{k=1}^m$, where $t_k' = t_k + \Delta t$, and the timestep, $\Delta t$, must be sufficiently small to resolve the highest frequencies in the dynamics.
As before, a snapshot may be the state of a system, such as a three-dimensional fluid velocity field sampled at a number of discretized locations that is reshaped into a high-dimensional column vector.
These snapshots are then arranged into two data matrices, $\bX$ and $\bX'$,
\begin{subequations}
\begin{align}
\bX &= \begin{bmatrix} \vline & \vline & & \vline \\
\bx(t_1) & \bx(t_2) & \cdots & \bx(t_m) \\
 \vline & \vline & & \vline
 \end{bmatrix}\\
 \bX' &= \begin{bmatrix} \vline & \vline & & \vline \\
\bx(t_1') & \bx(t_2') & \cdots & \bx(t_m') \\
 \vline & \vline & & \vline
 \end{bmatrix}.
\end{align}
\end{subequations}
If we assume uniform sampling in time, we will adopt the notation $\bx_k = \bx(k\Delta t)$.

The DMD algorithm seeks the leading spectral decomposition (i.e. eigenvalues and eigenvectors) of the  best-fit linear operator, $\bA$, that relates the two snapshot matrices in time by
\begin{align}
\bX' \approx \bA \bX. \label{eq:DMD-matrix}
\end{align}
The best-fit operator, $\bA$, then establishes a linear dynamical system that best advances snapshot measurements forward in time. If we assume uniform sampling in time, this becomes
\begin{align}
\bx_{k+1} \approx \bA \bx_k.\label{Eq:DMD:Propagator}
\end{align}
Mathematically, the best-fit operator $\bA$ is defined as
\begin{align}
\bA = \argmin_{\bA} \|\bX' - \bA \bX\|_F = \bX'\bX^\dagger\label{Eq:DMD:Definition}
\end{align}
where $\|\cdot\|_F$ is the Frobenius norm and $^\dagger$ denotes the Moore-Penrose pseudo-inverse\index{pseudo-inverse}.
The matrix $\bA$ is an operator that advances the measurements in $\bx$ forward in time.  It is often helpful to convert the eigenvalues of this discrete-time operator into continuous time, resulting in eigenvalues $\lambda=\mu+i\omega$. 

\REV{Alternative and better approaches are available~\cite{chen2012jns,jovanovic2014pof,askham2017arxiv} to the exact DMD algorithm. Bagheri~\cite{bagheri2014} first highlighted that DMD is particularly sensitive to the effects of noisy data, with systematic biases introduced to the eigenvalue distribution~\cite{duke2012error,bagheri2013jfm,dawson2016ef,hemati2017tcfd}. For example, when additive white noise is present in the measurements of an $n$-dimensional system with $m$ snapshots, the bias in exact DMD will be the dominant component of DMD error whenever the signal-to-noise ratio exceeds $\sqrt{n/m}$, implying that the effects of noise cannot always be mitigated by increasing the number of snapshots~\cite{dawson2016ef}.} As a result, a number of methods have been introduced to stabilize performance, including total least-squares DMD~\cite{hemati2017tcfd}, forward-backward DMD~\cite{dawson2016ef}, variational DMD~\cite{azencot2019consistent}, subspace DMD~\cite{takeishi2017}, time-delay embedded DMD~\cite{brunton2017natcomm} and robust DMD methods~\cite{askham2017arxiv,Scherl2020prf}.

However, the \emph{optimized} DMD algorithm of Askham and Kutz~\cite{askham2017arxiv}, which uses a variable projection method~\cite{askhamVariableProjectionMethods2018} for nonlinear least squares, provides the best performance of any algorithm currently available. This is not surprising given that it actually is constructed to both generalize and optimally satisfy the DMD problem formulation. In opt-DMD, the data matrix, $\mathbf{X}$, may be reconstructed as
\begin{eqnarray}
    \bX \approx  
    \underbrace{\left[ \! \begin{array}{ccc} | & & | \\ \boldsymbol{\phi}_1 & 
    \!\!\cdots\!\! & \boldsymbol{\phi}_r \\ | & & | \end{array} \! \right]}_{\bPhi} 
    \underbrace{\left[ \! \begin{array}{ccc} b_1 &  & \\ & \!\!\ddots\!\! & \\ & & b_r  \end{array} \!\right]}_{\mathrm{diag}(\bb)} 
    \underbrace{\left[ \! \begin{array}{ccc} e^{\lambda_1 t_1} & \!\cdots\! & e^{\lambda_1 t_m} \\
    \vdots & \!\ddots\! & \vdots \\ e^{\lambda_r t_1} & \!\cdots\! & e^{\lambda_r t_m} \end{array} \! \right]}_{{\bf T}(\boldsymbol{\lambda})},
\label{eq:dmd_opt}
\end{eqnarray}
where the  $i^{th}$ eigenmode, $\boldsymbol{\phi}_i$, has a corresponding mode amplitude $b_i$ and eigenvalue $\lambda_i$. The \REV{opt-DMD algorithm directly solves the exponential time dynamics fitting problem},
\begin{equation}
       \min_{ \boldsymbol{\lambda}, \bPhi_{\bf b} } \| \bX -   \bPhi_{\bf b} {\bf T}(\boldsymbol{\lambda}) \|_F,
\end{equation}
where \REV{modes and amplitudes are combined into $\bPhi_{\bf b} = \bPhi\mathrm{diag}(\bf b)$}. This has been shown to provide a superior decomposition due to its ability to optimally suppress bias. \REV{Unlike exact DMD and its variants, opt-DMD also computes an optimization contemporaneously across snapshots, eliminating the need for evenly timed samples.} The disadvantage of optimized DMD is that one must solve a nonlinear, nonconvex optimization problem. \REV{This is both computationally more expensive than exact DMD and the solution, by construction, only guarantees a local minimal to the optimization. In practice, opt-DMD is a relatively lightweight algorithm that has demonstrated itself to be numerically performant~\cite{askham2017arxiv}, relegating exact DMD and its variants like fbDMD to scenarios where low-precision is acceptable and low computational complexity is paramount.}

\section{Results and Analysis}

\begin{figure}
\begin{overpic}[width = 1.05\textwidth,trim = 0 120 0 80, clip]{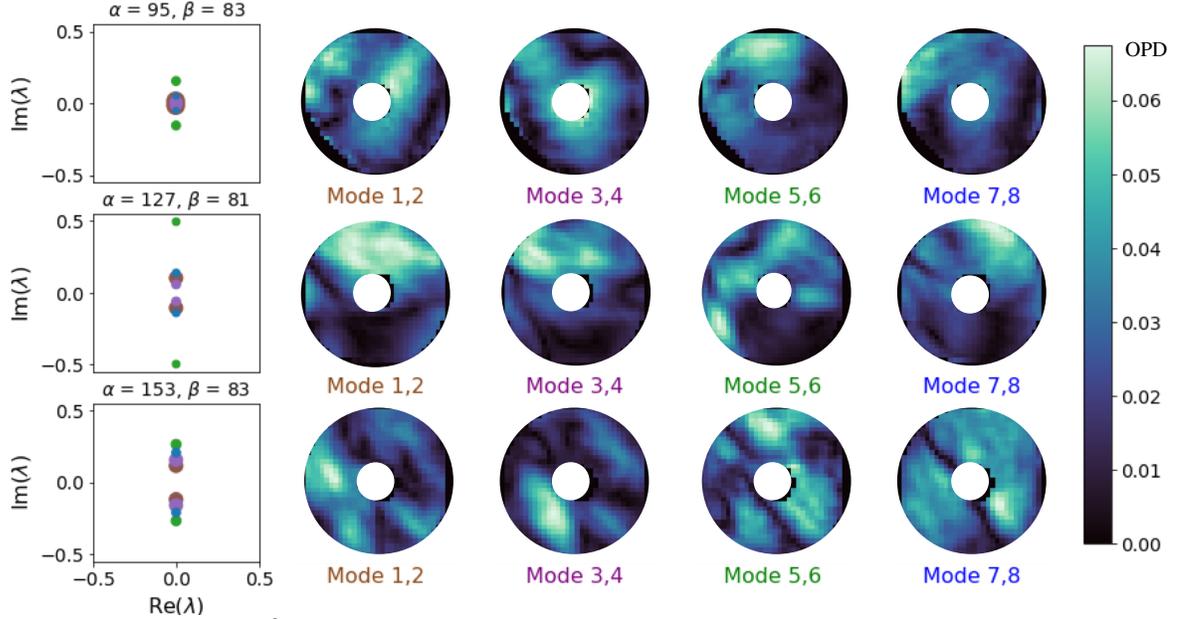}
\put(91.5,42){\small $^\text{OPD}$}
\end{overpic}   
\caption{Experiments with $\beta \approx 80^\circ$ for various $\alpha$, all in degrees. \REV{Each row depicts the truncated eigenvalue spectrum and first eight modes of the OPD for beam direction $(\alpha,\beta)$ as indicated above the eigenvalue plot. As seen across all trials, opt-DMD determines completely imaginary eigenvalues, as seen on the leftmost column of plots. These long-lasted modes enable long-time prediction. Mode labels are colored to match the associated eigenvalue in the eigenvalue plot.} Note that even modes are not displayed but are the complex conjugates of the odd modes.}
\label{fig:optdmd1} 
\end{figure}

\begin{figure}
\begin{overpic}[width = 1.05\textwidth,trim = 0 120 0 80, clip]{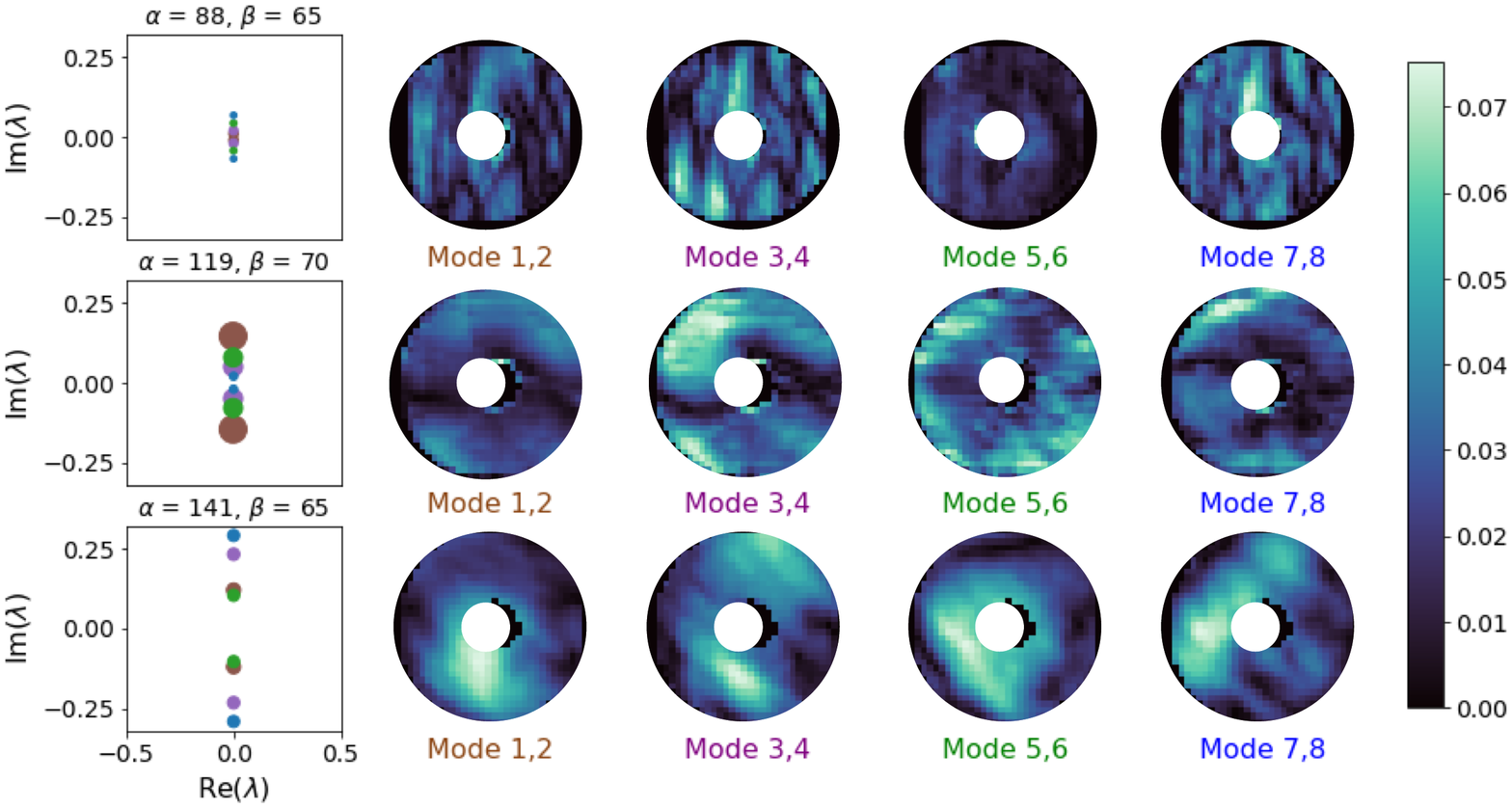}
\put(91.5,42){\small $^\text{OPD}$}
\end{overpic}
\caption{Experiments with $\beta \approx 65^\circ$ for various $\alpha$. \REV{Compare with Figures \ref{fig:optdmd1} and \ref{fig:optdmd3}.} Each row depicts the truncated eigenvalue spectrum and first eight modes of the OPD for beam direction $(\alpha,\beta)$ as indicated above the eigenvalue plot.}
\label{fig:optdmd2} 
\end{figure}

\begin{figure}
\begin{overpic}[width = 1.05\textwidth,trim = 0 120 0 80, clip]{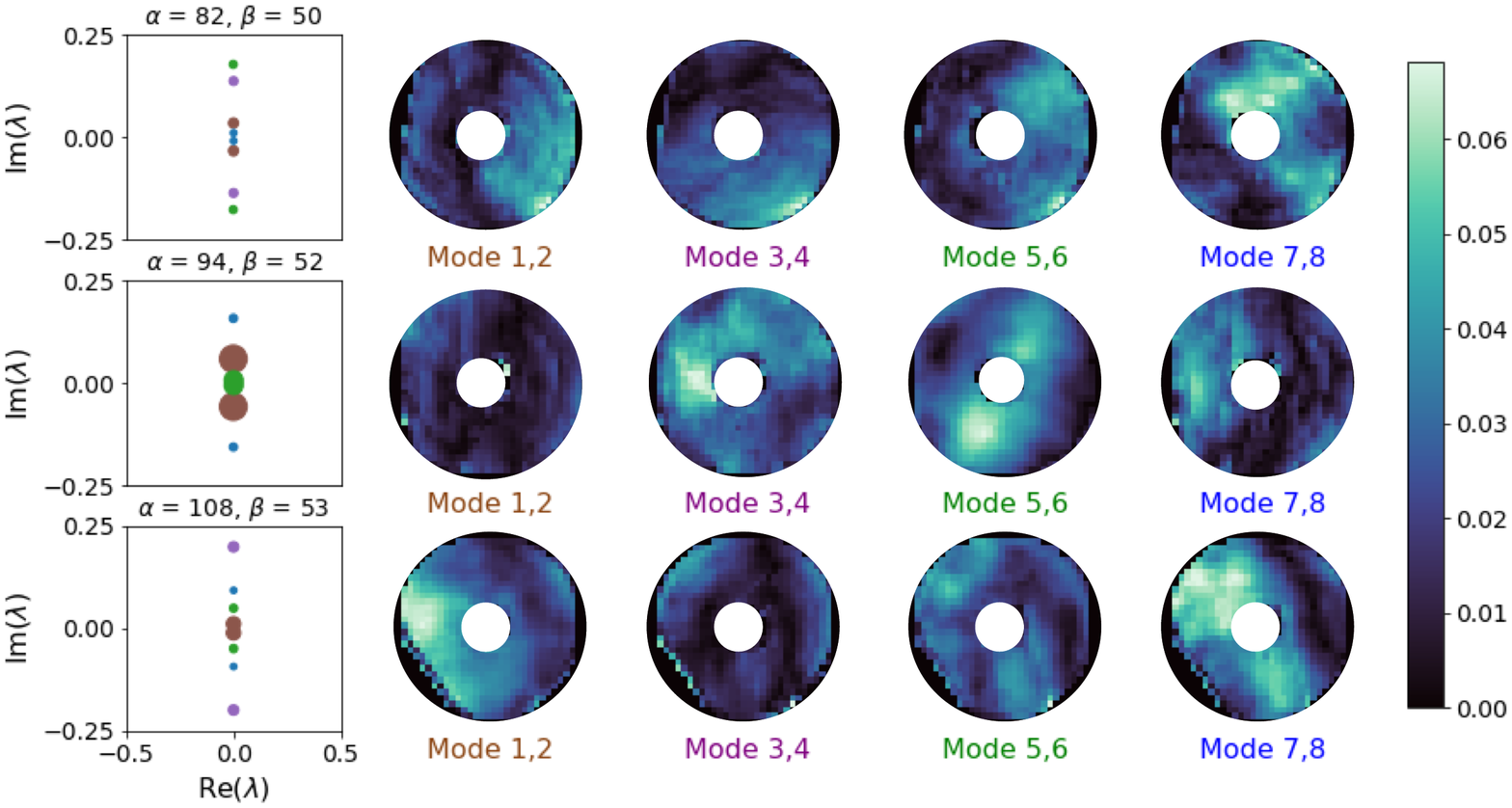}
\put(91.5,42){\small $^\text{OPD}$}
\end{overpic}
\caption{Experiments with $\beta \approx 50^\circ$ for various $\alpha$. \REV{Compare with Figures \ref{fig:optdmd1} and \ref{fig:optdmd2}.} Each row depicts the truncated eigenvalue spectrum and first eight modes of the OPD for beam direction $(\alpha,\beta)$ as indicated above the eigenvalue plot.}
\label{fig:optdmd3} 
\end{figure}

Figures \ref{fig:optdmd1}-\ref{fig:optdmd3} show the result of an opt-DMD analysis for a total of nine different turret angles $(\alpha, \beta)$: $(95^\circ, 83^\circ)$, $(127^\circ, 81^\circ)$, and $(153^\circ, 83^\circ)$ in Figure \ref{fig:optdmd1}; $(88^\circ, 65^\circ)$, $(119^\circ, 70^\circ)$, and $(141^\circ, 65^\circ)$ in Figure \ref{fig:optdmd2}; and $(82^\circ, 50^\circ)$, $(94^\circ, 52^\circ)$, and $(108^\circ, 53^\circ)$ in Figure \ref{fig:optdmd3}. \REV{These beam directions parameters were chosen to cover a variety of points along the AAOL-T's hemispherical turret. Grouping these trials by inclination angle, $\beta$, allows us to better compare look-back angles $\alpha$, as seen in the figures.}

Each row \REV{of Figures \ref{fig:optdmd1}-\ref{fig:optdmd3}} represents an individual $(\alpha, \beta)$ data set's opt-DMD analysis, showing the dominant eight eigenvalues and \REV{the corresponding modes} 1, 3, 5, and 7 from which the even numbered modes may be inferred as complex conjugates. Note in all cases, the eigenvalue spectrum is completely de-biased, lying along the imaginary axis. This is the critical takeaway of opt-DMD: with nearly perfect imaginary eigenvalues, the presented modes \REV{and their exponential time dynamics experience little time decay, allowing for arbitrarily long-lasting forecasts.}

\begin{figure}
\begin{overpic}[width = 1\textwidth,trim = 0 12 0 0, clip]{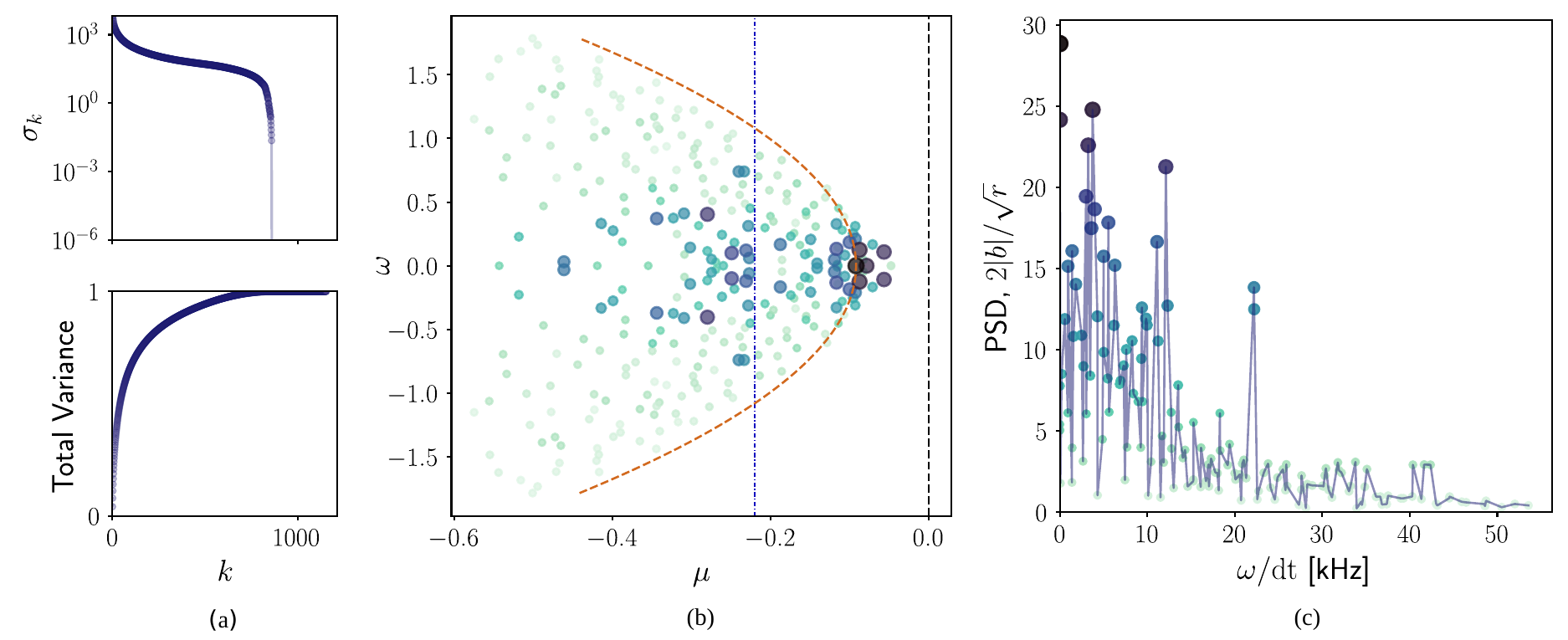}
\put(-1,5){(a)} \put(23,5){(b)} \put(62,5){(c)}
\end{overpic}
\caption{Demonstration of bias in the exact DMD algorithm for a beam direction with $\alpha=153^\circ, \beta=83^\circ$.  The bias produces a half-life decay for the forecast on the order of a hundred microseconds. \textbf{(a)}  SVD and cumulative energy of singular values. \textbf{(b)} Continuous-time eigenvalue spectrum. The orange dashed parabola which forms an envelope around the DMD eigenvalues is characteristic of noisy bias. In the opt-DMD spectra this curve becomes a vertical line. The blue dash-dotted line represents a cutoff to the left of which exist modes whose half-life exceeds  the mean half-life. \REV{Color and size matches eigenvalues to corresponding points in the next figure, (c), and the power spectral density's vertical axis.} \textbf{(c)} One-sided power spectrum of the DMD modes.}
\label{fig:exactdmd} 
\end{figure}

To compare with the precision of opt-DMD, we consider in Figure \ref{fig:exactdmd} an exact DMD analysis of the $(\alpha=153^\circ, \beta=83^\circ)$ dataset. As shown by the turret geometry Figure \ref{fig:turret}, this angle is roughly along the mid-line of the turret with a high look-back angle, pointing into the turbulent region prone to aero-optical effects but just outside regions with prominent horn vortices. The singular value spectrum and corresponding cumulative energy plots in Figure \ref{fig:exactdmd}{\color{blue}a} suggest an \REV{optimal} rank truncation $r=k=296$,~\cite{gavish2014arxiv} which is an overwhelming amount of modal detail to retain.

Figure \ref{fig:exactdmd}{\color{blue}b} shows the continuous-time eigenvalue spectrum of the system at the given rank truncation. The parabolic envelope $\mu(\omega)=-0.11\omega^2 - 0.09$ of the continuous-time eigenvalues ought to be compared with the spectrum of opt-DMD in Figures \ref{fig:optdmd1}-\ref{fig:optdmd3}, whose eigenvalues lie on the imaginary axis. \REV{The deformed envelope seen in Figure \ref{fig:exactdmd}{\color{blue}b}} is consistent with weak noise on self-sustaining oscillating flow fields.~\cite{bagheri2014}. While truncating the exact DMD analysis at a lower rank may produce modes closer to the imaginary axis, a parabolic envelope remains and the performance of opt-DMD remains superior by construction. 

\REV{Computing the modal half-life gives us a window into the shortcomings of exact DMD. The mean half-life is found to be}
\begin{equation}
    \langle t_{1/2}\rangle = \frac{1}{r}\sum_{j=1}^r \frac{-\log(2)\Delta t}{\mu_j} = \SI{104}{\micro\second}
\end{equation}
and the amplitude-weighted mean half-life is
\begin{equation}
    \langle t_{1/2}^b\rangle = \frac{1}{\sum_{i=1}^r |b_i|} \sum_{j=1}^r \frac{-|b_j|\log(2)\Delta t }{\mu_j} = \SI{138}{\micro\second}.
\end{equation}
\REV{The modal half-life indicates a window of opportunity for a predictor to interact with an AO control loop. For the particular example in Figure \ref{fig:exactdmd}, the modal half-lives individually spanned a range from $\SI{50}{} - \SI{500}{\micro\second}$. When examining all 23 trials for various $(\alpha,\beta)$, we discovered the mean half-lives were consistently on the order of a hundred microseconds. With these decay timescales, pertinent coherent turbulent structures become treated as transient effects, diminishing the ability of DMD to forecast the dominant spatiotemporal structures on a long horizon.} Figure \ref{fig:exactdmd}{\color{blue}c} characterizes the power spectrum of the exact DMD modes. Note that many powerful modes have lower than average half-lives, further compromising the ability of exact DMD to forecast turbulent flow dynamics.

\section{Conclusion}

Data-driven methods are becoming increasingly important to model complex spatio-temporal systems whose evolution dynamics are not well known or only characterized by time-series measurements. In the case of aero-optic interactions, modeling the induced turbulent wake from a turret is exceptionally challenging. Unless dynamics are characterized in an appropriate manner, the wavefront aberrations cannot be corrected in the AO system. We proposed a data-driven algorithmic architecture which aims to model the aero-optic interactions in an adaptive and real-time manner. Specifically, we introduced the opt-DMD algorithm to produce an unbiased modal analysis of the AAOL-T dataset which captures the wavefront aberrations induced by a turbulent flow around a turret. The imaginary-valued eigenspectrum of the opt-DMD operator permits longer forecasting in an AO loop when compared to an exact DMD algorithm. \REV{Exact DMD suffers from modal decay rates due to the real components introduced in the spectrum, whereas opt-DMD modes display no decay rate and permit forecasting as long as the model remains physically meaningful with respect to the environment.  Indeed, traditional DMD algorithms have forecasting horizons which decay on the order of hundreds of microseconds whereas opt-DMD by construction allows for forecasting on scales required for control algorithms for AO corrections.}

Further studies ought to assess the performance of opt-DMD on turret geometries beyond hemispherical, as well as compare opt-DMD's forecasting ability to existing aero-optical predictors that rely on time-invariant POD modes for dimensionality reduction or neural network architectures~\cite{burnsRobustModificationPredictive2016}. \REV{Importantly, opt-DMD's minimal bias as well as its freedom in sampling variable time steps make it a promising predictor for aero-optical phenomena.}



\subsection* {Acknowledgments}  SLB acknowledges funding support from the Air Force Office of Scientific Research (AFOSR FA9550-19-1-0386) and the Army Research Office (ARO W911NF-19-1-0045). 
\\
\\
Portions of this manuscript were previously published in SPIE Proceedings.~\cite{kutz_physics-informed_2021}

\subsection*{Disclosures} Approved for public release; distribution is unlimited. Public Affairs release approval \#AFRL-2021-3106.


\bibliography{refs}   
\bibliographystyle{spiejour}   


\section{Biographies}

\vspace{2ex}\noindent\textbf{Shervin Sahba} is a Physics Ph.D. Candidate at the University of Washington. He received his M.S. degree in Applied Mathematics from the University of Washington in 2019,  M.S. in Physics from San Francisco State University in 2017, and B.A. in Psychology and B.S. in Business Management from the University of Rhode Island in 2005. His research explores photonic systems and physics-informed machine learning.

\vspace{2ex}\noindent\textbf{Diya Sashidhar} is a Ph.D. Candidate in Applied Mathematics at the University of Washington. She received her B.S. in Applied Mathematics from North Carolina State University in 2017. Her research focuses on time series forecasting and uncertainty quantification, machine learning, and data-driven modeling.

\vspace{2ex}\noindent\textbf{Christopher C. Wilcox} is an Electrical Engineer at the US Air Force Research Laboratory. He earned his B.S. in Electrical Engineering at the New Mexico Institute of Mining and Technology, M.S. in Electrical and Computer Engineering from University of New Mexico, and Ph.D. in Engineering at University of New Mexico in 2009. He is a Research Fellow at the US Naval Postgraduate School Adaptive Optics Center of Excellence in National Security and SPIE Fellow.

\vspace{2ex}\noindent\textbf{Austin McDaniel} received his B.A. degree in Mathematics from the University of Pennsylvania and his Ph.D. degree in Applied Mathematics from the University of Arizona. He is a mathematician at the US Air Force Research Laboratory.

\vspace{2ex}\noindent\textbf{Steven L. Brunton} is a Professor of Mechanical Engineering at the University of Washington, an Adjunct Professor of Applied Mathematics and Computer Science, and a Data Science Fellow at the eScience Institute. Steve received his B.S. in mathematics from Caltech in 2006 and the Ph.D. in mechanical and aerospace engineering from Princeton in 2012. His research combines machine learning with dynamical systems to model and control systems in fluid dynamics, biolocomotion, optics, energy, and manufacturing.

\vspace{2ex}\noindent\textbf{J. Nathan Kutz} received his B.S. degree in Physics and Mathematics from the University of Washington, Seattle, WA, USA, in 1990, and Ph.D. in Applied Mathematics from Northwestern University, Evanston, IL, USA, in 1994. He is currently Director of the AI Institute in Dynamic Systems, a Professor of Applied Mathematics, an Adjunct Professor of Physics, Mechanical Engineering, and Electrical and Computer Engineering, and a Senior Data Science Fellow of the eScience Institute, University of Washington. 

\end{spacing}
\end{document}